Spin photonics on chip based on a twinning crystal metamaterial

Yan Li[†], Jingbo Sun[†, *], Yongzheng Wen, Xiaoyu Xiong and Ji Zhou*

State Key Laboratory of New Ceramics and Fine Processing, School of Materials Science and

Engineering, Tsinghua University, Beijing 100084, China.

[†] These authors contributed equally to this work.

[*] To whom it should be addressed: jingbosun@tsinghua.eud.cn, zhouji@tsinghua.edu.cn

**Abstract:** Two-dimensional photonic circuits with high capacity are essential for a wide range of applications in next-generation photonic information technology and optoelectronics. Here we demonstrate a multi-channel spin-dependent photonic device based on a twinning crystal metamaterial. The structural symmetry and material symmetry of the twinning crystal metamaterial enable a total of 4 channels carrying different transverse spins because of the spin-momentum locking. The orientation of the anisotropy controls the propagation direction of each signal, and the rotation of the E-field with respect to energy flow determines the spin characteristics during input/output coupling. Leveraging this mechanism, the spin of an incident beam can be maintained during propagation on-chip and then delivered back into the free space, offering a new scheme for metamaterial-based spin-controlled nano-photonic applications.

The rotation of electric and magnetic fields in a light beam are relevant to a wide range of optical effects. In classical electromagnetics these are called polarization states, but from a quantum mechanical point of view they can also be understood as the intrinsic spin of a photon or spin angular momentum. Engineering with this property has shown significant potential in chip-size photonic information technology and optoelectronics, as well as for the study and engineering of light-matter interactions[1-11]. Photon spin can be easily generated using anisotropic materials through the introduction of a phase difference between different linear polarizations, resulting in circular polarizations. However, splitting beams of opposite spins is more difficult, as the photonic spin Hall effect which causes this is rather weak in bulk material[12-16]. One solution to enhance this effect is to flatten the bulk material into two dimensions, eschewing volumetric fields and dealing instead with optical surface waves[5, 9-11, 17]. Dyakonov surface waves (DSWs)[18-24] which are supported at the interface between isotropic and anisotropic media are an emerging candidate for this application. As laterally confined surface waves, their fields rotate around an axis that is orthogonal to the wave vector, giving rise to transverse spin[25] whose direction is determined by the anisotropy of the interface[19]. In this case, the spin-dependent propagation of the DSW has a built-in directionality that is totally independent of the coupling structure, and thus can be used to transmit coded signals[26].

In this work, we experimentally showed an ultra-compact multi-channel photonic circuit with spin-dependent input, propagation, and output that is based on a crystal twinning metamaterial with symmetric anisotropy. Incident beams with left/right circular polarization can be coupled into DSW modes through the twin boundary directly. Inside the twinning crystal metamaterial, the propagating DSW modes includes 4 channels with different transverse spins and built-in directionality which are determined by the orientations of the mirrored twin grains. By properly designing the output coupling structure, the characteristic spin carried by the incident beam can be successfully preserved, being transferred into transverse spins in the surface wave channels and then delivered back into free space. Therefore, the crystal twinning metamaterial itself serves as an integrated photonic chip.

The entire twinning crystal is comprised of two mirrored twin grains located in xy-plane with the twin boundary along x-axis as shown in Fig. 1a. Both grains are implemented through a hyperbolic metamaterial made of silver grooves where the optical axis (OA) of the hyperbolic metamaterial is normal to the grooves. The twinning metamaterial has structural symmetry with respect to the xz-plane and material symmetry (*e.g.*, isotropy/anisotropy/isotropy) across the xy-plane, thereby resulting in 4 different spin states. If we assume that the wave vector of the DSW ($k_{DSW}$) in the right grain is along +y direction at the interface in between the air (isotropic) and the metamaterial (anisotropic), and given an orientation angle of $\varphi$ with respect to OA, then E-field at the interface is a linear combination of the transverse electric (TE) and transverse magnetic (TM) modes[18]:

$$\text{TE:} \quad E_{TE} = A_{TE} (1 \quad 0 \quad 0) \exp[i(k_{DSW} y - \omega t)] \tag{1a}$$

$$\text{TM:} \quad E_{TM} = A_{TM} (0 \quad -ik_{DSW} \quad \frac{k_{DSW}}{k_0}) \exp[i(k_{DSW} y - \omega t)] \tag{1b}$$

where $k_0$ is the wave vector in the free space. According to Equation (1), the component along $k_{DSW}$ (**$E_y$**) has $\pi/2$ lag compared to the E-field normal to $k_{DSW}$ (**$E_r$**=**$E_x$**+**$E_z$**) at the interface between air (isotropic) and the metamaterial (anisotropic), and thus a transverse spin of $\sigma^+$, which is ruled by the spin-momentum locking[27]. At the lower interface in between the metamaterial (anisotropic) and the glass substrate (isotropic), the transverse spin is reversed: $\sigma^-$, because of the spin-momentum locking[27]. Due to the structural symmetry of the twinning metamaterial, the DSWs at the left grain have opposite propagation momentum (-$k_{DSW}$), and thus the transverse spins of opposite handedness. As a result, the twinning metamaterial effectively produces 4 types of interfaces, one in each of the 4 quadrants of the yz-plane, which can be viewed as 4 separate channels (Ch1, 2, 3, and 4). These channels can support different transverse spin states, which are plotted in the inset in Fig. 1a, in Movie 1, and are also summarized in the 3rd and 4th rows in Table I.

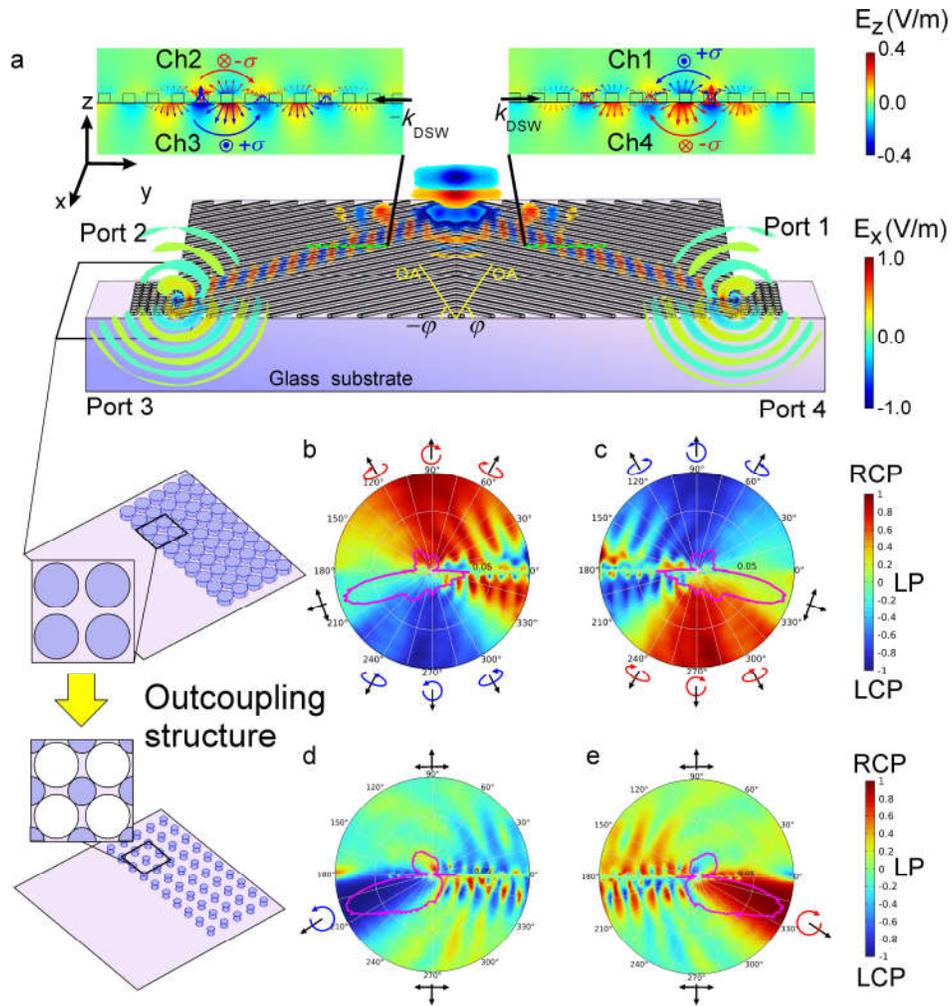

Figure 1 The schematic of the crystal twinning metamaterial-based spin-photonic circuit. (a) E-field distributions under a linear x-polarized (LP) illumination at the twin boundary. DSW modes are excited with wave vectors of ± $k_{DSW}$ along y-axis. Simulated $E_x$ fields are shown for the incident field, surface waves, and the output fields. The inset shows simulated $E_z$-fields as well as the transverse spin vectors of the E-fields for each of the 4 channels (Ch1, 2, 3, and 4) inside the DSW modes along the green dashed lines. (b) and (c) show the radiation field distribution with the coupling structure of large cylinders. The color scales show the spin density along the radial direction and the pink curves show the amplitude of the E field. As shown by the polarization states, the output can also be sorted into 4

ports. (d) and (e) show the radiation field distribution with the coupling structure of small cylinders. The main radiations is in LCP at Port 3 in (d) and in RCP at Port 4 in (e).

Now we use the twinning metamaterial as a photonic circuit with spin-dependent input, propagation, and output. First, the twin boundary itself is a one-dimensional defect on the metamaterial film, which also serves as an input coupler with deep subwavelength size in the yz-plane. As shown in Fig. 1a, the twin boundary is along x-axis. When a light beam is incident onto the twin boundary, the excited scattering wave vectors are mainly perpendicular to the twin boundary and thus one can expect the coupled DSWs' wave vectors $\pm k_{DSW}$ along the y-axis. In the simulation, DSWs with counterpropagating wave vectors $\pm k_{DSW}$ are generated in both grains with normally-incident, x-polarized light. This result is exactly the same as the simulated result when using a uniform twinning crystal with effective dielectric parameters (Methods and Fig. S1). Directionality shows up once we switch to circularly-polarized incident light due to interactions between the spin and the orientation of the anisotropy. Illuminating with a left circularly-polarized (LCP) beam, only one DSW mode with $k_{DSW}$ along the +y (Ch1 and 4) was generated in the right grain (Fig. S2), whereas right circularly-polarized (RCP) incident light resulted in a DSW mode in the left grain (Ch2 and 3) with $-k_{DSW}$ (Fig. S3).

Here we find that if we check the rotation of the E field with respect to the Poynting vector, the spins of the light in free space and the surface wave mode may show a very high consistency. In free space, LCP and RCP show counter clockwise rotation (CCW) and clockwise rotation (CW) of the E field regarding to either wave vector or the Poynting vector. At the interfaces, the wave vector and the Poynting vectors are not along with each other due the anisotropy, and then the transvers spin $\sigma^+$ in

Ch1 also shows CCW of the E field while the σ- of Ch2 show clockwise (CW) rotation regarding to the Poynting vectors. Under such a definition, LCP in free space and $σ^+$ of the DSW show the same spin characters of CCW, while the RCP and $σ^-$ are the same for the E field rotation of CW. Simultaneously, in each grain, Ch1 vs Ch4 and Ch2 vs Ch3 have opposite spins, and these will finally determine the polarization states at the 4 output ports (Movie 2 and 3).

At the output side, we use an array of silver cylinders for output coupling, which scatters isotropically in order to maintain the spin property of the DSW mode as much as possible. Fig. 1b shows the simulated polarization state distributions coupled into free space from the left grain. The E-field rotation of the output coupling to the upper space (z>0) is CW, which is consistent with an RCP input and the transverse spin in Ch2 at the upper interface (Movie 2). Along the +z direction, it is almost perfectly RCP. Similarly, the output coupling to the lower space (z<0) is CCW and is LCP along –z direction, which in consistent with the $σ^-$ in Ch3 at the lower interface (Movie 2). According to this rule, at the right side, we find that the upward output is LCP, since it originates from a combination of LCP incidence and propagation in Ch1, while the coupling downwards is RCP, as shown in Fig. 1c and Movie 3. In this case, the 4 channels' spin-dependent signals in the twinning metamaterial can be out-coupled to different directions through 4 ports (Port 1, 2, 3, and 4). Their spin relations are summarized in the 5th and 6th rows in Table I. From these results we can draw a very intuitive conclusion. The spin characters of the inputs, surface wave propagations, and the outputs are well conserved as either circular polarizations (longitudinal spin) or transverse spin. The rotations of the E-field are unchanged with respect to the Poynting vector throughout the entire process.

The pink curves in Fig. 1b and c show the radiation intensity of at the left and right output ports, respectively. Despite the pure LCP and RCP output light being parallel to the z-axis, the maximum intensities of the output beams are linearly-polarized and are not along the z-axis but rather at angles of approximately 195º and 345º on the left and right sides (both of which are in the –z half-space). To create the spin-dependence in the main output, we use small cylinders which are evolved from the complementary pattern of the original output port cylinder array (Fig. S4). Fig. 1d and e show the polarization states, which are almost inverted distributions compared to those in Fig. 1b and c. Despite the inversion, the intensity of the output radiation remains the same, with the main output directions of all beams being in the -z half-space. In this case, the two maximum outputs are loaded with spins, and thus replicate the transverse spins at the metamaterial/glass interface, carrying LCP at Port 3 and RCP at Port 4, respectively.

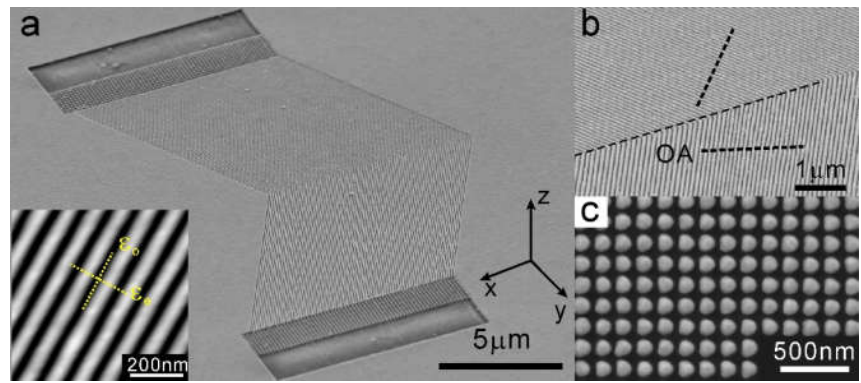

Figure 2 Scanning electron microscope images of the twinning metamaterial based photonic circuit by FIB. (a) The overall image of the twinning metamaterial with output coupler. The inset shows the silver grooves with a period of 80 nm and 50% duty cycle; (b) the twin boundary between the two mirrored symmetrical grains. The dashed lines shows the OAs of the two grains; (c) The cylinder array output coupler.

The entire pattern, including grooves and the output structures, was milled by Focused Ion Beam (FIB) on a 60nm thick silver film with glass substrate in one-step, as shown in Fig. 2. Based on the analysis above, an incident beam is applied to the twin boundary from the air side, and output signals are collected with a high NA objective at Port 3 and 4, where the intensities greatest. We also note that Port 3 is at the right side and Port 4 is at the left side of the twinning metamaterial in Fig. 3 and 4 because the photos are taken from the glass side.

Table I. Spin-dependent input, propagation, and output.

| Input | | RCP | LCP |
|---|---|---|---|
| Twinning metamaterial | | Left grain | Right grain |
| Interfaces | Air/metamaterial | Ch2: $\sigma-$ | Ch1: $\sigma+$ |
| | metamaterial/glass | Ch3: $\sigma+$ | Ch4: $\sigma-$ |
| Output | $z>0$ | Port 2: RCP | Port 1: LCP |
| | $z<0$ | Port 3: LCP | Port 4: RCP |

To test the performance of such a prototype spin encoding device based on a crystal twinning metamaterial, we first used an x-polarized laser at 808nm as the input. As shown in Fig. 3a, we observed outputs from the two grains simultaneously, coupled from the DSWs propagating along the surfaces of the metamaterials. These show no directionality, which is similar to the results of previous work[9] using SPP coupling under LP incident light. However, the difference is that our coupling structure is the twin boundary which does not have any phase modulation functionality or chirality. The propagation directions of the LCP and RCP components in the LP incidence were determined by the orientations of the metamaterials. Once we switched the polarization state of the incident beam from LP into LCP, Port 4 showed emissions, while the other output at Port 3 should be off. However, there is still some intensity in Fig. 3b and d, which we will later show to be scattered fields and not real output from Port 3. Reciprocally, when the incident polarization was switched to RCP, the output signal is only from Port 3, as shown in Fig. 3c and d. These results demonstrate that LCP incident fields are coupled into the DSW mode in Ch1 and 4 in the right grain ($\varphi_0=60°$), and RCP incident fields are coupled into the DSW mode of metamaterial grain on the left (Ch 2 and 3) with $\varphi_0=-60°$.

In Fig. 3b, there is still some residual intensity at Port 3, which can be attributed to random stray scatterings since the intensity profile is totally different from the one at Port 3 in Fig. 3a. Here we performed a similarity analysis on the output patterns among Fig. 3a, b and c by using the HASS method. Zoomed-in regions which are around the output ports (outlined by white boxes in Fig. 3a-c) are plotted in Fig. 3d. The intensity profiles inside the pink box of Ports 4 from Fig. 3a and b match each other very well, as shown in Fig. 3e, which proves that the DSW coupled from LCP incidence was the same as the LCP component of the LP incidence, since this LCP light can only be coupled into Ch1 and 4. As shown in Fig. 3f, the output patterns at the Port 3 under LP and RCP incidence are quite similar, implying that RCP incidence is directed toward Ch2 and 3. All these spin-dependent behaviors at the input coupling are consistent with the results in Table I.

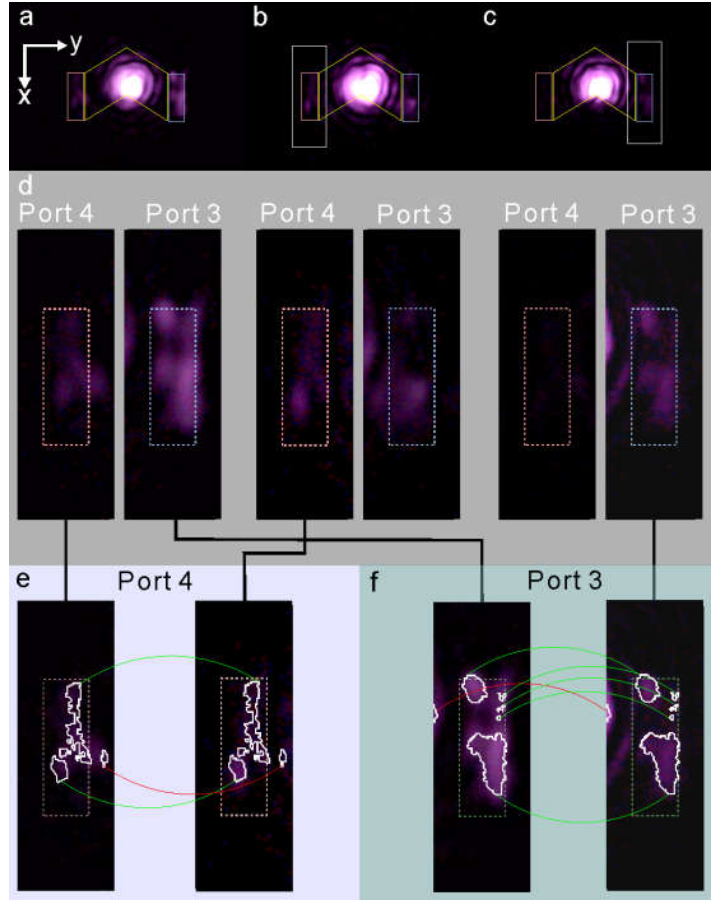

Figure 3 Spin-dependent input coupling behavior. (a) LP incidence; (b) RCP incidence; (c) LCP incidence; (d) Magnified output regions; (e) Similarity analysis of Port 4 from (a) and (b) based on HASS method. (f) Similarity analysis of Port 3 from (a) and (c) based on HASS method. Since all the photos are taken from the glass side, Port 4 is on the left and Port 3 is on the right side in these photos.

We proved that the DSWs inherit the certain spins from the incident beam and the propagation direction was completely determined by the orientation of the metamaterial. However, whether the carried transverse spin can be further delivered into free space and measured in the far field was still uncertain. As shown in Fig. 4a, the outputs at Ports 3 and 4 are from the LCP and RCP components of the LP incident wave, respectively. Since we are collecting the output from the glass side, where the radiation intensity is at its maximum, the polarization states actually follow the transverse spins of Ch3 and Ch4 at the metamaterial/glass interface, which are opposite to those of the incident field. Here we use a quarter waveplate and a linear polarizer as an analyzer to determine the polarization state of the output signals from each grain. At the beginning, both the transmission axis of the linear polarizer and the fast axis of the quarter waveplate are oriented horizontally to the optical table. Then we rotate the linear polarizer by 45º in CCW, we observed that the output at Port 4 is kept while the other one is much darker, as shown in Fig. 4b. This means the current output from Port 4 is RCP. Then we rotate the polarizer by 90º so that the polarizer is now 45º degrees lag of the fast axis, an inverse phenomenon is observed where the beam from Port 4 is filtered and the output at Port 3 appears, which can be shown to be LCP, as shown in Fig. 4c.

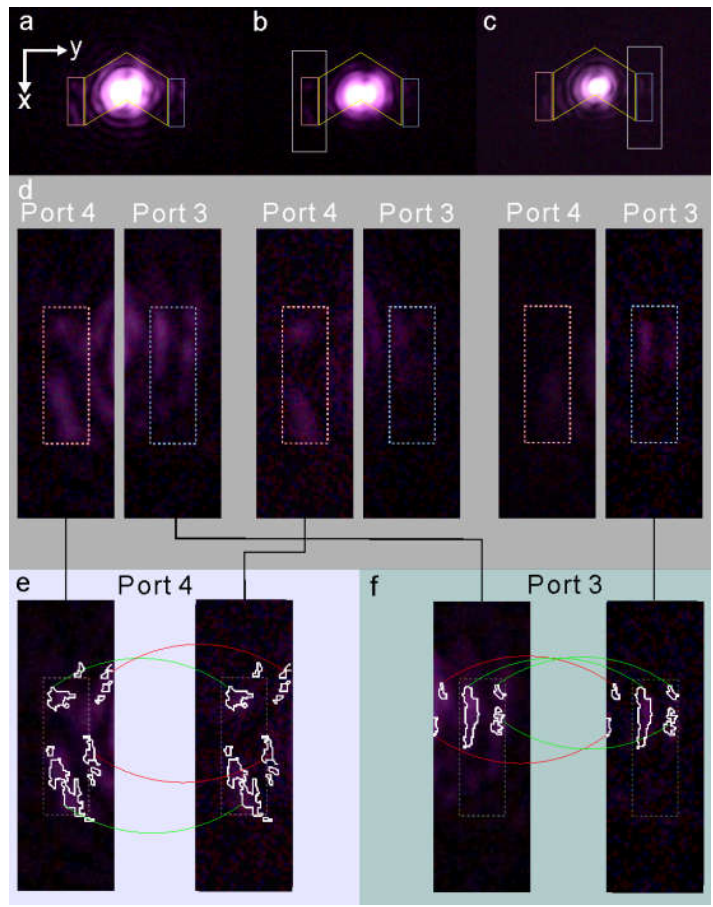

Figure 4 Spin-dependent output coupling behavior. (a) LP incidence; (b) LCP analyzer; (c) RCP analyzer; (d) Magnified output regions; (e) Similarity analysis of Port 4 from (a) and (b) based on the HASS method. (f) Similarity analysis of Port 3 from (a) and (c) based on the HASS method. Since all the photos are taken from the glass side, Port 4 is on the left and Port 3 is on the right side in these photos.

In order to avoid noise due to scattered light, we also performed similarity analyses on the output beam patterns with and without the quarter waveplate and polarizer. Regions around the output structure outlined by the white box are zoomed-in and shown in Fig. 4d. The output pattern's profile at Port 4 in Fig. 4b shows a rather high similarity to the Port 4 under the LP incidence in Fig. 4a, which implies the RCP output coupled from the DSW. Meanwhile, the output pattern from the Port 3 in Fig. 4b is totally different from that in the LP case, which might be due to scattered fields but not the real output. Reciprocally, the output pattern at Port 3 in Fig. 4c shows almost the same pattern but weakened compared to the Port 3 output in the LP case while the output at Port 4 in Fig. 4c has been filtered out entirely after the linear polarizer. Therefore, these results demonstrate that the spin carried by DSWs can be well maintained during out-coupling and collected in the free space.

In conclusion, we have demonstrated a spin-dependent multi-channel photonic circuit by using a twinning crystal metamaterial. Based on the intrinsic symmetries of the twinning crystal metamaterial, different transverse spin states can be spatially separated and out-coupled into free space in different directions. The polarization states of the incident wave from free space, transverse spins of the surface modes, and the polarization states of the output back to the free space can be linked together by the rotation of the E-field with respect to the Poynting vector. As a consequence, the characteristic spin carried by the incident beam can be well maintained throughout the process of the input coupling, propagation and output coupling. Such a twinning crystal metamaterial provides an integrated scheme for the deployment of next-generation high-capacity spin photonics technologies.

**Methods**

Modelling.

The twinning metamaterial contain two symmetrically oriented HMMs made of silver grooves, where silver's filling ratio is 50%. The period of the groove is 80 nm and the thickness of the metamaterial is 60 nm. The extraordinary permittivity $\varepsilon_e$ and ordinary permittivity $\varepsilon_o$ of HMM are the permittivities normal and parallel to the grooves, respectively. The permittivity of silver is -31.69+0.4i[28]. By using effective medium theory, at the wavelength of 808nm, $\varepsilon_e$ is 2.06 and $\varepsilon_o$ is -15.34-0.2i. According to our previous work, DSW modes may exist in such a HMM in a rather wide angle range 24~90°, based on its hyperbolic anisotropy ($\varepsilon_e$, $\varepsilon_o$). The wave vectors of the DSW modes are always along the y-axis, and therefore the angles between $k_{DSW}$ and the OA are $\pm\varphi_0=60°$, which fall in the allowed angle range. The numerical simulations of the twinning crystal metamaterial using actual grooves structures and a twinning crystal of uniform slab with effective parameters ($\varepsilon_o$, $\varepsilon_e$) were carried out using COMSOL Multiphysics. By comparing these two results which are almost exactly the same with each other with respect to $k_{DSW}$, energy flow, and the transverse spin, we are sure that the silver groove structure is working as a hyperbolic medium, supporting DSWs but not the grooved metallic surface for the SPPs.

Characterization

The optical setup to characterize this multi-channel chip is plotted in Fig. S5 in the supplementary material. An 808 nm laser is coupled onto the sample through a 20× objective. At the other side of the sample, the output beams are collected through a 100× objective and then imaged by a CCD camera. The CCD camera at the incident side is used to determine the location of the twin boundary at the sample. In front of the sample, a linear polarizer and a quarter waveplate are used to control the polarization states of the incoming beam. At the output side, the other set of linear polarizer and quarter waveplate are working as the analyzer to test the handedness of the circular polarization of the output beam.

**Acknowledgements**

This work was supported by the National Key R&D Program of China (Grant No. 2022YFB3806000), National Natural Science Foundation of China under Grant Nos. 11974203, 12074420, 11704216 and 52072203.

Competing interests

The authors declare no competing financial interest.

**Author Contributions**

J.S. proposed the idea and made the design. J.S. performed the numerical simulations. Y.L. did the fabrication of the samples and performed the optical characterization of the sample. J.S. wrote the paper. Y.W. contributed to the discussions. J.S. and J.Z. supervised this work.

**Supplementary**

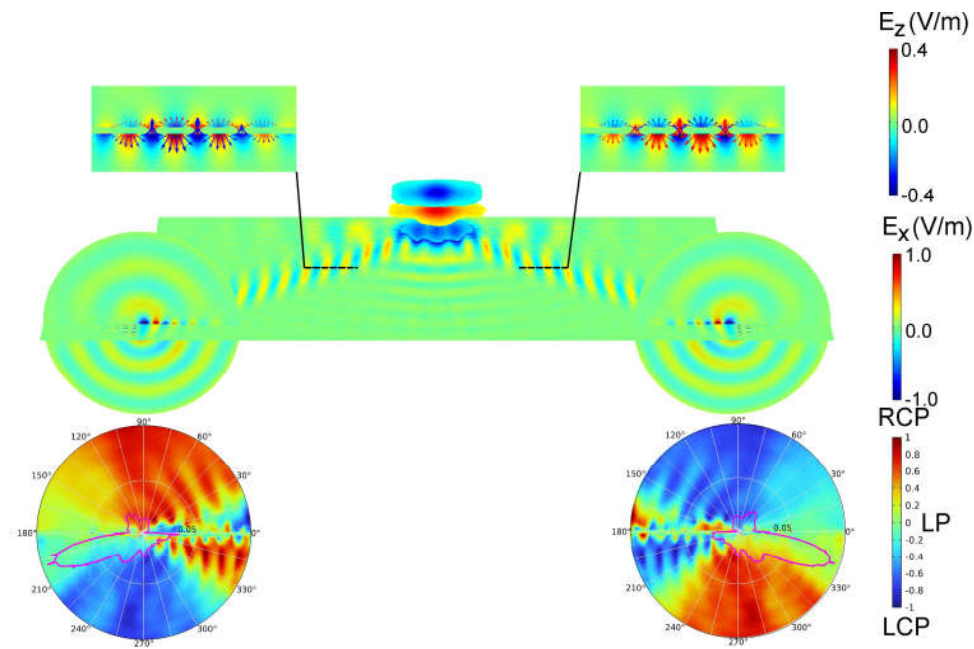

Fig. S1 Simulations on a hyperbolic twinning crystal of effective medium.

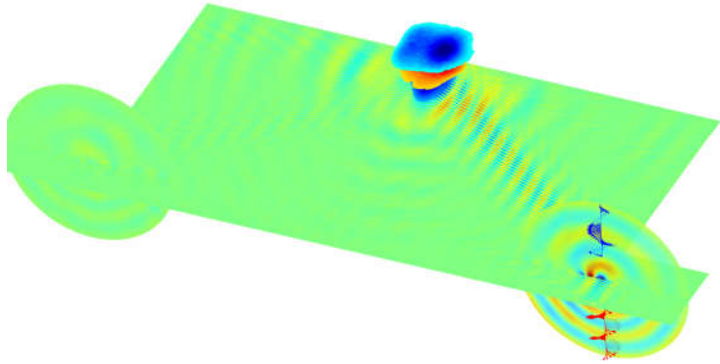

Fig. S2 DSW excited by left circularly polarized incidence.

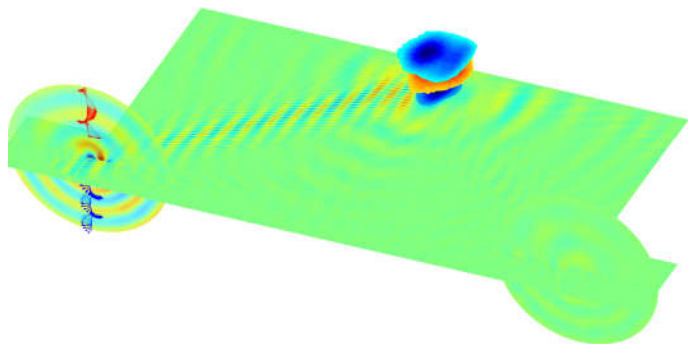

Fig. S3 DSW excited by right circularly polarized incidence.

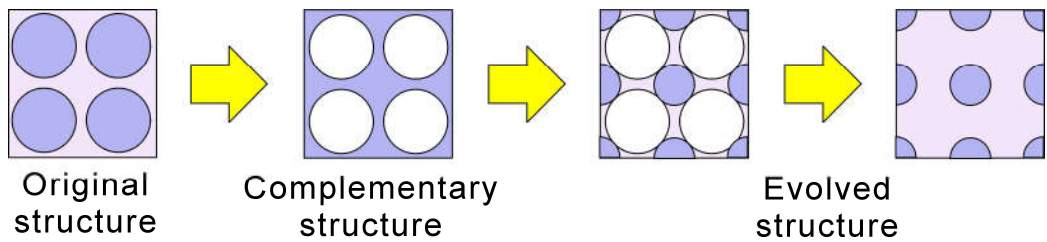

Fig. S4 Output coupling structure.

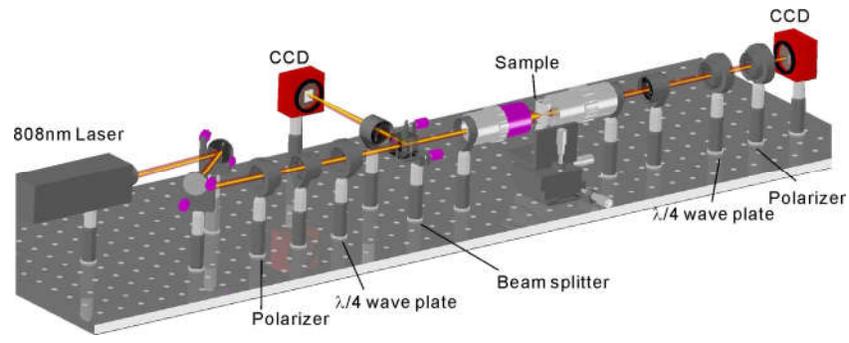

Fig. S5 Characterization setup.